\documentclass[manuscript, nonacm]{acmart}
\AtBeginDocument{%
  }

\setcopyright{none}
\renewcommand\footnotetextcopyrightpermission[1]{} % Removes bottom text

\begin{document}

\title{Hidden Technical Debt in Generative (GenUI) and Malleable User Interfaces}

\author{Besjon Cifliku}
\email{besjon.cifliku@cais-research.de}
\orcid{0009-0007-5081-9531}
\affiliation{%
  \institution{Center for Advanced Internet Studies (CAIS)}
  \city{Bochum}
  \country{Germany}
}

\renewcommand{\shortauthors}{Besjon Cifliku}

\begin{abstract}
Malleable software can profoundly change how users interact with digital content, enabling non-experts to create their own customized tools. However, the practical adoption of GenUI systems faces several barriers, which I unpack in this paper, including a lack of adaptable data formats, ``old'' security protocols, and gaps in users' cognitive and creative skills for building their own interfaces. I advocate new evaluation strategies and scientific methods to measure the impact of malleable software in user studies, document usage patterns, and ensure their practical adoption.
\end{abstract}



\maketitle

\section*{Paper accepted to the Workshop ``What Does Generative UI Mean for HCI Practice?'' at CHI 2026, ACM CHI Conference on Human Factors in Computing Systems 2026 -- Barcelona, April 15, 2026. }
\vspace{1.3em}

\section{Motivation \& Research Vision}

Digital interfaces have mainly remained static for years~\cite{BeaudouinLafon2004DesignInteractions}, requiring users to click through windows, drag-and-drop files, search, and share data across multiple applications. Without technical expertise, users rely on prebuilt software settings, leaving app development centralized among developers~\cite{Litt2025MalleableSoftware}. Allan Key’s vision of users ``changing the tools'' as they edit documents challenges this narrative by offering customizable, adaptable software~\cite{Kay1984OpeningHood}.

MacLean et al. acknowledged the impossibility of designing systems that meet all users' needs and situations~\cite{MacLean90s}. However, GenUI prioritizes users' ability to create their own interfaces and edit the underlying data structure. Rather than a developer explicitly coding and enforcing the behavior, the system synthesizes interface components on demand, often in response to user prompts or inferred based on usage patterns, context, and user needs~\cite{Litt2025MalleableSoftware}. Prior work has examined users' perspectives on GenUI~\cite{DefiningGenUIInterfaces, XiangChen2025}, focusing on the development of technical frameworks to support such systems~\cite{GenerativeMalleableUI, MaesGenUIne2024, min2026gradualgenerationuserinterfaces}. These efforts include designing overview and detailed interface catalogs~\cite{MalleableOverviewDetail}, elucidating user requirements for generative systems~\cite{XiangChen2025}, and developing a new AI instruction framework to bridge the gap between developers, designers, and end users while establishing design patterns for malleable interfaces~\cite{MeridianMalleableUI}. Additional efforts have been directed toward developing novel AI-UI protocols to integrate generative capabilities into UIs~\cite{a2ui2025, mcp_apps2025}.

Despite the promises of accessibility and interactivity, malleable interfaces are not widely adopted~\cite{Yucel2025ComplexityFlexibleInterfaces, AgainstGenUI}. The applications remain primarily as research prototypes. In this paper, I highlight issues that prevent GenUI tools from being deployed in production. I categorize the limitations into: \textbf{1) }data sharing complexity, \textbf{2)} non-adaptive security protocols, and \textbf{3)} skills gap and cognitive overload. I label these challenges as technical debt~\cite{Cunningham92TechnicalDebt} stemming from GenUI adoption, compounded by an incomplete understanding of dynamic software patterns when creating malleable interfaces and by unclear guidelines how to integrate for integrating them. 

\section{Hidden Debt: The Gaps Towards Generative UI}

When using GenUI, drafting the first application is quick~\cite{XiangChen2025}, with plenty of available tools~\cite{Bolt2024, OpenAI2024Codex, Anthropic2024ClaudeCode, Google2024AIStudio, Lovable2024}. However, underlying issues emerge when users need additional context and manual effort to reach a fully working version~\cite{XiangChen2025}. Moreover, most non-technical users are unaware of underlying issues such as database management, security limitations, and architectural scaling. 

\subsection{How Malleable Is Your Data?}
Software consists of predefined, hard-coded data models that determine the system's behavior, the state of display, and how it handles errors gracefully. If we rely on current infrastructure, most software communicates with different providers to sustain their services (e.g., Google, Microsoft, Meta). Users will inevitably need to exchange information with external parties. Malleable software might violate established data transfer protocols, making it hard to adapt to new, incompatible formats. Furthermore, users' modifications are difficult to maintain as the underlying application or data schema evolves~\cite{Nazmus2020MITEmbodiedMathematics, BeaudouinLafon2004DesignInteractions}. Researchers proposed task-driven data models that ensure consistency~\cite{GenerativeMalleableUI}, and provide transparent records of abstractions to ease inspection by end users~\cite{Hayatpur2025ShapesOfAbstraction}. For instance, a collaboration library called Automerge demonstrates how to sync JSON documents from multiple users, eliciting a foundation for malleable data exchange~\cite{Kleppmann2018Automerge, Litt2025MalleableSoftware}. We could rethink the design of adaptable data protocols by implementing a centralized, shared collaboration database that is updated after each new iteration. This database needs to handle secure information exchange while incorporating robust anonymizing and encryption constraints to ensure user privacy is not compromised.

\subsection{Can We Generate Adaptive Security Protocols Too?} 
The aforementioned data issues highlight a broader need for emerging security protocols and privacy-preserving policies. Malleable interfaces, by design, empower users to repurpose and customize software to align with their specific needs~\cite{Litt2025MalleableSoftware}. However, this flexibility introduces several security risks, even when applications run in sandbox environments. User modifications may cause unpredictable updates, leaving their systems vulnerable to previously unknown security exploits. Without anti-mitigation strategies against cyber threats, users may inadvertently expose themselves to malicious or unintended user interfaces that compromise system security and leave them vulnerable to novel-gen UI social engineering phishing attacks (see \cite{clawdbot_infostealers_2026}). To address these issues, the research field must develop adaptive testing methodologies and new security principles. These approaches should prioritize mechanisms that record the outputs of generative processes, including the conditions under which they were produced, facilitating future understanding and reproduction.

\subsection{What Users Really Need?}

The most insidious form of tech deb in GenUI is the skills gap and limited understanding of how such systems work. In prior work~\cite{Cifliku2025LBW, Cifliku2026}, I explored the integration of AI-supported automation tools in journalism, revealing persistent challenges in user engagement and skill acquisition. One recurring theme was journalists' perception of insufficient creativity to leverage these technologies effectively and to design their own tools. Another concern is the lack of knowledge about what they can do with AI, what is possible, and the limitations of such systems~\cite{Jones2022AIEverywhereAndNowhere} (see also ~\cite{Litt2025MalleableSoftware}). Users may find that the learning curve for these mechanisms offers no immediate reward when cognitive overload is high, discouraging non-technical users from adopting such systems~\cite{Litt2025MalleableSoftware, MacLean90s}. Designing one's own tools requires a higher level of abstraction~\cite{Victor2011LadderOfAbstraction, Girardin2026SoftwareGetsPersonal}, resulting in cognitive burden for the user~\cite{Nazmus2020MITEmbodiedMathematics, GenerativeMalleableUI}. The literacy gap contributes to this behavior, compounded by a lack of training, limited expertise, and inadequate computational skills. In addition, the opaque prompt sensitivity and the volatile, non-deterministic nature of AI-generated outputs~\cite{Razavi2025PromptSensitivity, Azrabzadeh2025PromptSensitivity} create an uninviting (``hostile''~\cite{Victor2011LadderOfAbstraction}) ecosystem for non-technical end-users who would need to interpret, control, and validate the resulting artifacts~\cite{GenerativeMalleableUI}. Ultimately, the key question stands out: How can we encourage users to embrace the role of creators rather than passive consumers of pre-packaged software~\cite{Litt2025MalleableSoftware}? This shift requires a systemic and epistemic efforts to cultivate the skills needed to navigate and shape GenUI tools safely, without compromising users' autonomy or agency. 

\section{Looking Ahead}

This workshop is essential to shape GenUI’s future in HCI, rethinking interaction paradigms~\cite{BeaudouinLafon2004DesignInteractions} and empowering users to adapt their digital tools. As we strive to create systems that write and repurpose themselves, addressing these hidden debt demands new research practices and perhaps new theoretical frameworks. From an HCI perspective, we must balance output control with user agency. Applying version-control principles, such as branching and merging, across media types could let users explore tool variations safely, enabling traceability and interpretability~\cite{Litt2025MalleableSoftware}. These tools could be toggled on demand, granting users granular administrative power without overwhelming casual users. 

HCI research must redefine new evaluation metrics~\cite{Ledo2018Evaluation} with regard to interpretability of the system, learnability, and the affected users' agency and autonomy~\cite{BennetHowAgencyAutonomyHCI2023}. In theory, a software system that adapts to user actions could proportionally reduce the user's cognitive load as long as they continue to engage with it. Still, it might also introduce new, unanticipated edge cases that were not considered in the initial research plan. The open research question remains: \textbf{1. How to track and audit a continuously changing software system without hampering users' agency while making this process transparent to users?} The problem lies in the control needed to promote security without endangering users' privacy and autonomy; and \textbf{2. How would we reproduce the findings of a research experiment when the interface continuously changes substantially in response to newly released protocols and AI's capabilities?} The System Usability Scale (SUS)~\cite{SUS} results might not tell the whole story; usability can fluctuate depending on the components generated, even with well-defined component catalogs. In conclusion, I argue that achieving malleable software in practice requires a fundamental shift in how we design software~\cite{BeaudouinLafon2004DesignInteractions} and, most importantly, in how the community researches these domains in the HCI context. 

\bibliographystyle{ACM-Reference-Format}
\bibliography{bibliography}

@techreport{Kay1984OpeningHood,
  author       = {Alan C. Kay},
  title        = {Opening the Hood of a Word Processor},
  institution  = {Self-published / Worrydream Refs},
  year         = {1984},
  number       = {Draft},
  url = {https://worrydream.com/refs/Kay_1984_-_Opening_the_Hood_of_a_Word_Processor.pdf},
  note         = {Draft working paper (distributed for comments only)}, 
}

@online{Litt2025MalleableSoftware,
  author       = {Geoffrey Litt and Josh Horowitz and Peter van Hardenberg and Todd Matthews},
  title        = {Malleable Software: Restoring User Agency in a World of Locked-Down Apps},
  howpublished = {Ink \& Switch, June 2025},
  url          = {https://www.inkandswitch.com/essay/malleable-software/},
  note         = {Accessed: 2026-02-08},
  year = {2025}
}

@inproceedings{MacLean90s,
author = {MacLean, Allan and Carter, Kathleen and L\"{o}vstrand, Lennart and Moran, Thomas},
title = {User-tailorable systems: pressing the issues with buttons},
year = {1990},
isbn = {0201509326},
publisher = {Association for Computing Machinery},
address = {New York, NY, USA},
url = {https://doi.org/10.1145/97243.97271},
doi = {10.1145/97243.97271},
booktitle = {Proceedings of the SIGCHI Conference on Human Factors in Computing Systems},
pages = {175–182},
numpages = {8},
location = {Seattle, Washington, USA},
series = {CHI '90}
}

@inbook{DefiningGenUIInterfaces,
author = {Lee, Kyungho},
title = {Towards a Working Definition of Designing Generative User Interfaces},
year = {2025},
isbn = {9798400714863},
publisher = {Association for Computing Machinery},
address = {New York, NY, USA},
url = {https://doi.org/10.1145/3715668.3736365},
booktitle = {Companion Publication of the 2025 ACM Designing Interactive Systems Conference},
pages = {489–495},
numpages = {7}
}

@inproceedings{GenerativeMalleableUI,
author = {Cao, Yining and Jiang, Peiling and Xia, Haijun},
title = {Generative and Malleable User Interfaces with Generative and Evolving Task-Driven Data Model},
year = {2025},
isbn = {9798400713941},
publisher = {Association for Computing Machinery},
address = {New York, NY, USA},
url = {https://doi.org/10.1145/3706598.3713285},
doi = {10.1145/3706598.3713285},
booktitle = {Proceedings of the 2025 CHI Conference on Human Factors in Computing Systems},
articleno = {686},
numpages = {20},
location = {
},
series = {CHI '25}
}

@inproceedings{MeridianMalleableUI,
author = {Min, Bryan and Xia, Haijun},
title = {Meridian: A Design Framework for Malleable Overview-Detail Interfaces},
year = {2025},
isbn = {9798400720376},
publisher = {Association for Computing Machinery},
address = {New York, NY, USA},
url = {https://doi.org/10.1145/3746059.3747654},
doi = {10.1145/3746059.3747654},
booktitle = {Proceedings of the 38th Annual ACM Symposium on User Interface Software and Technology},
articleno = {200},
numpages = {14},
location = {
},
series = {UIST '25}
}

@inproceedings{MalleableOverviewDetail,
author = {Min, Bryan and Chen, Allen and Cao, Yining and Xia, Haijun},
title = {Malleable Overview-Detail Interfaces},
year = {2025},
isbn = {9798400713941},
publisher = {Association for Computing Machinery},
address = {New York, NY, USA},
url = {https://doi.org/10.1145/3706598.3714164},
doi = {10.1145/3706598.3714164},
booktitle = {Proceedings of the 2025 CHI Conference on Human Factors in Computing Systems},
articleno = {688},
numpages = {25},
location = {
},
series = {CHI '25}
}

@online{Yucel2025ComplexityFlexibleInterfaces,
  author       = {Mustafa Y{\"u}cel},
  title        = {The Complexity of Creating Flexible Interfaces: Why Flexibility Needs Clear Intent},
  year         = {2025},
  month        = may,
  howpublished = {Medium},
  url          = {https://compeng.medium.com/the-complexity-of-creating-flexible-interfaces-why-flexibility-needs-clear-intent-dae5ec02a2c7},
  note         = {Accessed: 2026-02-08},
}

@article{Cunningham92TechnicalDebt,
author = {Cunningham, Ward},
title = {The WyCash portfolio management system},
year = {1992},
issue_date = {April 1993},
publisher = {Association for Computing Machinery},
address = {New York, NY, USA},
volume = {4},
number = {2},
issn = {1055-6400},
url = {https://doi.org/10.1145/157710.157715},
doi = {10.1145/157710.157715},
journal = {SIGPLAN OOPS Mess.},
month = dec,
pages = {29–30},
numpages = {2}
}

@online{OpenAI2024Codex,
  author       = {{OpenAI}},
  year         = {2026},
  title        = {ChatGPT Codex},
  howpublished = {OpenAI},
  url          = {https://chatgpt.com/codex},
  note         = {Accessed: 2026-02-08},
}

@online{Anthropic2024ClaudeCode,
  author       = {{Anthropic}},
  title        = {Claude Code: Overview},
  howpublished = {Anthropic documentation},
  url          = {https://code.claude.com/docs/en/overview},
  note         = {Accessed: 2026-02-08},
}

@misc{Google2024AIStudio,
  author       = {{Google}},
  year         = {2026},
  title        = {Google AI Studio},
  howpublished = {Google},
  url          = {https://aistudio.google.com/},
  note         = {Accessed: 2026-02-08},
}

@misc{Lovable2024,
  author       = {{Lovable}},
  title        = {Lovable: Build Software with AI},
  year         = {2026},
  howpublished = {Lovable},
  url          = {https://lovable.dev/},
  note         = {Accessed: 2026-02-08},
}

@misc{Bolt2024,
  author       = {{Bolt}},
  title        = {Bolt},
  howpublished = {Bolt},
  url          = {https://bolt.new/},
  note         = {Accessed: 2026-02-08},
}

@inproceedings{XiangChen2025,
author = {Chen, Xiang 'Anthony and Knearem, Tiffany and Li, Yang},
title = {The GenUI Study: Exploring the Design of Generative UI Tools to Support UX Practitioners and Beyond},
year = {2025},
isbn = {9798400714856},
publisher = {Association for Computing Machinery},
address = {New York, NY, USA},
url = {https://doi.org/10.1145/3715336.3735780},
doi = {10.1145/3715336.3735780},
booktitle = {Proceedings of the 2025 ACM Designing Interactive Systems Conference},
pages = {1179–1196},
numpages = {18},
location = {
},
series = {DIS '25}
}

@phdthesis{Nazmus2020MITEmbodiedMathematics,
  author       = {Saquib, Nazmus},
  title        = {Embodied mathematics by interactive sketching},
  school       = {Massachusetts Institute of Technology},
  department   = {Program in Media Arts and Sciences, School of Architecture and Planning},
  year         = {2020},
  publisher    = {Massachusetts Institute of Technology},
  url          = {https://hdl.handle.net/1721.1/129275},
  note         = {Ph.D. thesis. Cataloged from student-submitted PDF. Includes bibliographical references (pp. 189--197)},
}

@misc{Girardin2026SoftwareGetsPersonal,
  author       = {Fabien Girardin},
  title        = {Software Gets Personal: An Introduction},
  year         = {2026},
  month        = jan,
  howpublished = {Medium (Pr\'{o}ximo Presents)},
  url          = {https://medium.com/pr\'{o}ximo-presents/software-gets-personal-an-introduction-1175c7f1edbd},
  note         = {13 min read; Accessed: 2026-02-08},
}

@inproceedings{Ledo2018Evaluation,
  author       = {David Ledo and Steven Houben and Jo Vermeulen and Nicolai Marquardt and Lora Oehlberg and Saul Greenberg},
  title        = {{Evaluation Strategies for HCI Toolkit Research}},
  booktitle    = {Proceedings of the 2018 CHI Conference on Human Factors in Computing Systems},
  series       = {CHI '18},
  year         = {2018},
  month        = apr,
  pages        = {36:1--36:17},
  publisher    = {ACM},
  address      = {New York, NY, USA},
  doi          = {10.1145/3173574.3173610},
  url          = {https://doi.org/10.1145/3173574.3173610},
}

@misc{Victor2011LadderOfAbstraction,
  author       = {Bret Victor},
  title        = {Up and Down the Ladder of Abstraction: A Systematic Approach to Interactive Visualization},
  year         = {2011},
  howpublished = {Worrydream.com},
  url          = {https://worrydream.com/LadderOfAbstraction/},
  note         = {Interactive essay; Accessed: 2026-02-08},
}

@inproceedings{Cifliku2025LBW,
author = {Cifliku, Besjon and Heuer, Hendrik},
title = {"This could save us months of work" - Use Cases of AI and Automation Support in Investigative Journalism},
year = {2025},
isbn = {9798400713958},
publisher = {Association for Computing Machinery},
address = {New York, NY, USA},
url = {https://doi.org/10.1145/3706599.3719856},
doi = {10.1145/3706599.3719856},
booktitle = {Proceedings of the Extended Abstracts of the CHI Conference on Human Factors in Computing Systems},
articleno = {29},
numpages = {8},
location = {
},
series = {CHI EA '25}
}

@inproceedings{Cifliku2026,
author = {Cifliku, Besjon and Heuer, Hendrik},
title = {They Think AI Can Do More Than It Actually Can: Practices, Challenges, \& Opportunities of AI-Supported Reporting In Local Journalism},
year = {2026},
publisher = {Association for Computing Machinery},
address = {New York, NY, USA},
url = {https://doi.org/10.1145/3772318.3791130},
doi = {10.1145/3772318.3791130},
booktitle = {Proceedings of the 2026 CHI Conference on Human Factors in Computing Systems},
numpages = {20},
series = {CHI'26}
}

@article{Jones2022AIEverywhereAndNowhere,
  author    = {Bronwyn Jones and Rhianne Jones and Ewa Luger},
  title     = {AI ‘Everywhere and Nowhere’: Addressing the AI Intelligibility Problem in Public Service Journalism},
  journal   = {Digital Journalism},
  year      = {2022},
  volume    = {10},
  number    = {10},
  pages     = {1731--1755},
  doi       = {10.1080/21670811.2022.2145328},
  url       = {https://doi.org/10.1080/21670811.2022.2145328},
}

@inproceedings{BennetHowAgencyAutonomyHCI2023,
author = {Bennett, Dan and Metatla, Oussama and Roudaut, Anne and Mekler, Elisa D.},
title = {How does HCI Understand Human Agency and Autonomy?},
year = {2023},
isbn = {9781450394215},
publisher = {Association for Computing Machinery},
address = {New York, NY, USA},
url = {https://doi.org/10.1145/3544548.3580651},
doi = {10.1145/3544548.3580651},
booktitle = {Proceedings of the 2023 CHI Conference on Human Factors in Computing Systems},
articleno = {375},
numpages = {18},
location = {Hamburg, Germany},
series = {CHI '23}
}

@inproceedings{BeaudouinLafon2004DesignInteractions,
author = {Beaudouin-Lafon, Michel},
title = {Designing interaction, not interfaces},
year = {2004},
isbn = {1581138679},
publisher = {Association for Computing Machinery},
address = {New York, NY, USA},
url = {https://doi.org/10.1145/989863.989865},
doi = {10.1145/989863.989865},
booktitle = {Proceedings of the Working Conference on Advanced Visual Interfaces},
pages = {15–22},
numpages = {8},
location = {Gallipoli, Italy},
series = {AVI '04}
}

@inproceedings{Hayatpur2025ShapesOfAbstraction,
author = {Hayatpur, Devamardeep and Hempel, Brian and Lin, Richard and Xia, Haijun},
title = {The Shapes of Abstraction in Data Structure Diagrams},
year = {2025},
isbn = {9798400713941},
publisher = {Association for Computing Machinery},
address = {New York, NY, USA},
url = {https://doi.org/10.1145/3706598.3713723},
doi = {10.1145/3706598.3713723},
booktitle = {Proceedings of the 2025 CHI Conference on Human Factors in Computing Systems},
articleno = {883},
numpages = {12},
location = {
},
series = {CHI '25}
}

@inproceedings{Kleppmann2018Automerge,
  author       = {Martin Kleppmann and Alastair R. Beresford},
  title        = {{Automerge: Real-time data sync between edge devices}},
  booktitle    = {Proceedings of the 1st UK Mobile, Wearable and Ubiquitous Systems Research Symposium (MobiUK 2018)},
  year         = {2018},
  pages        = {101--105},
  url          = {https://mobiuk.org/abstract/S4-P5-Kleppmann-Automerge.pdf},
  note         = {Abstract; accessed 2026-02-08},
}

@inproceedings{Razavi2025PromptSensitivity,
author = {Razavi, Amirhossein and Soltangheis, Mina and Arabzadeh, Negar and Salamat, Sara and Zihayat, Morteza and Bagheri, Ebrahim},
title = {Benchmarking Prompt Sensitivity in Large Language Models},
year = {2025},
isbn = {978-3-031-88713-0},
publisher = {Springer-Verlag},
address = {Berlin, Heidelberg},
url = {https://doi.org/10.1007/978-3-031-88714-7_29},
doi = {10.1007/978-3-031-88714-7_29},
booktitle = {Advances in Information Retrieval: 47th European Conference on Information Retrieval, ECIR 2025, Lucca, Italy, April 6–10, 2025, Proceedings, Part III},
pages = {303–313},
numpages = {11},
location = {Lucca, Italy}
}

@inproceedings{Azrabzadeh2025PromptSensitivity,
author = {Arabzadeh, Negar and Clarke, Charles L.A.},
title = {A Human-AI Comparative Analysis of Prompt Sensitivity in LLM-Based Relevance Judgment},
year = {2025},
isbn = {9798400715921},
publisher = {Association for Computing Machinery},
address = {New York, NY, USA},
url = {https://doi.org/10.1145/3726302.3730159},
doi = {10.1145/3726302.3730159},
booktitle = {Proceedings of the 48th International ACM SIGIR Conference on Research and Development in Information Retrieval},
pages = {2784–2788},
numpages = {5},
location = {Padua, Italy},
series = {SIGIR '25}
}

@misc{a2ui2025,
  title        = {{A2UI: A Protocol for Agent-Driven Interfaces}},
  author       = {{Google}},
  howpublished = {\url{https://a2ui.org}},
  year         = {2025},
  note         = {Accessed: 2026-02-12},
  url          = {https://a2ui.org}
}

@online{mcp_apps2025,
  title  = {MCP Apps-Model Context Protocol Extensions Documentation},
  author = {Model Context Protocol},
  year   = {2025},
  url    = {https://modelcontextprotocol.io/docs/extensions/apps},
  urldate= {2026-02-12}
}

@misc{min2026gradualgenerationuserinterfaces,
      title={Gradual Generation of User Interfaces as a Design Method for Malleable Software}, 
      author={Bryan Min and Peiling Jiang and Zhicheng Huang and Haijun Xia},
      year={2026},
      eprint={2601.17975},
      archivePrefix={arXiv},
      primaryClass={cs.HC},
      url={https://arxiv.org/abs/2601.17975}, 
}

\appendix

\end{document}